\documentclass[aps,prd,twocolumn,10pt]{revtex4}
\usepackage{xcolor}
\definecolor{Gray}{gray}{0.85}
\definecolor{LightCyan}{rgb}{0.88,1,1}
\usepackage[colorlinks,linkcolor={red},citecolor={blue}]{hyperref}
\usepackage{eurosym}
\usepackage{graphicx}
\usepackage{amsmath}
\usepackage{amssymb}
\usepackage{amsfonts}
\usepackage{bm}
\usepackage{hhline}

\usepackage{multirow}

\def\be{\begin{equation}}
	\def\ee{\end{equation}}
\def\bea{\begin{eqnarray}}
	\def\eea{\end{eqnarray}}

\newcommand{\mc}[1]{\mathcal{#1}}

\newcommand{\f}[2]{\frac{#1}{#2}}

\begin{document}
\title{Cosmological implications of hybrid metric-Palatini $f(\mathcal{R},\mc{R}_{\mu\nu}\mc{R}^{\mu\nu})$ gravity}
\author{Shahab Shahidi}
\email{s.shahidi@du.ac.ir}
\author{Shiva Kayedi}
\email{sh.kayedi@std.du.ac.ir}
\affiliation{School of Physics, Damghan University, Damghan 36716-45667, Iran.}
\date{\today}
\begin{abstract}
	The hybrid metric-Palatini gravity with Lagrangian density $L =R+f(\mathcal{R},\mc{R}_{\mu\nu}\mc{R}^{\mu\nu})$ is considered, where $R$ is the metric Ricci scalar and $\mc{R}_{\mu\nu}$ is the Palatini Ricci tensor. Contrary to the standard hybrid metric-Palatini theory, because of the term $\mathcal{R}_{\mu\nu}\mathcal{R}^{\mu\nu}$ in the action, the model can not be analytically transformed to a scalar-tensor theory. However, on top of a maximally symmetric space-times like the FRW universe, there is a way to solve for a metric compatible connection which we will follow in this paper. The cosmological implications of the resulting model will then be fully considered. The best fit values of the model and cosmological parameters will be obtained by confronting the model with the recent observational data on the Hubble parameter. We will see that the observational data can be explained very good in this model, but, significant deviations from the standard $\Lambda$CDM model could be seen in derivatives of the Hubble parameter $q$, $j$ and $s$. We will perform a statefinder analysis for the model and show that its behavior differs from that of the $\Lambda$CDM model. Also, we will consider the recently proposed $Om$ diagnostics to categorize the dark energy type of the model and obtain the $\omega$-varying alternative of the model  that mimics the Hubble flow.
\end{abstract}
\maketitle
\section{Introduction}
Late-time observations of the type Ia supernovae \cite{super} suggests that the universe is experiencing an accelerated expanding phase at present times. This conclusion is also approved by the observations of the baryon acoustic oscillations \cite{BAO}. As a first attempt to explain the accelerated expansion of the universe, one can restore the cosmological constant proposed originally by Einstein \cite{firsteinstein}. This, together with the addition of cold dark matter to the energy budget of the universe, will form the standard 
$\Lambda$CDM model on which all observations are based \cite{LCDM}.

Despite that the $\Lambda$CDM model is theoretically simple, minimal and beautiful, it suffers from both observational and theoretical issues like the cosmological constant problem \cite{CCproblem}, the coincidence problem \cite{coniprob} and the $H_0$ and $\sigma_8$ tensions \cite{tensions}. These issues together with our theoretical curiosities, encourages us to explore for more advanced alternatives to Einstein general relativity which are collectively known as the generalized theories of gravity. 

The generalized theories of gravity can be achieved in several ways. The first possibility includes theories with modified gravity/geometry, like the $f(R)$ gravities \cite{fR}, higher dimensional models \cite{brane}, massive gravity theories \cite{massive}, Weyl geometric models \cite{weyl}, Finsler geometries \cite{finsler}, etc. Beside observational constraints, the main concern in modified gravity theories is the presence of ghost instabilities, which restricts the form of the final action \cite{ghost}.

The second possibility includes addition of more matter fields to the Einstein's theory which is known as modified matter gravities. This can be done by adding other fields like the scalar/vector theories \cite{scalarvector}, generalizing the matter Lagrangian itself to include non-standard functions of the matter Lagrangian, the trace of the energy momentum tensor, etc. \cite{generalmatter} or adding derivatives of the matter Lagrangian \cite{derivative}. The main problem in this category is to correctly define the energy momentum tensor \cite{secondderivativepaper}.

Another possibility to generalize the Einstein general relativity which makes more attention recently, includes coupling between matter and geometry sectors, generally called the matter-geometry coupling theories. More famous Lagrangians include $f(R,T)$ \cite{fRT}, $f(R,L_m)$ \cite{fRLm}, $f(R,T,R_{\mu\nu}T^{\mu\nu})$ \cite{fRTRT} and $f(R,T,L_m)$ \cite{fRTLm} models. There is however a debate that these types of theories can not be considered independent. They are either unstable or lie in the generalized matter category \cite{mukohyama}. The debate is replied in \cite{replytomukoohyama} where the authors explain what is special in these type of theories and why they are stable and independent.

One of the first generalizations of the Einstein's general relativity, that lies in the first category, is to consider an independent affine connection instead of the metric-driven Christoffel connection. The resulting theory, known as the Palatini model, is proved to be equivalent to the Einstein' theory if one considers the Einstein-Hilbert action as a starting point \cite{palatini}. However, one obtains a different theory for more general action \cite{generalPalatini}. The Palatini approach can then be though as an alternative viewpoint to all generalized theories of gravity \cite{palatinipapers}. Recently, another idea was developed that the action contains both viewpoints together \cite{hybridmetricpalatini}. In this hybrid metric-Palatini model, one write the action as
\begin{align}
	S=\int d^4x\sqrt{-g}\,\Big[\kappa^2 \big(R+\mathcal{R}\big)+L _m\Big],\nonumber
\end{align}
where the first Ricci scalar is constructed from the metric tensor and the connection is the Christoffel symbol, and the second Ricci scalar is constructed from the metric tensor and an independent affine connection. This can also be generalized to $R+f(\mathcal{R})$, where $f$ is an arbitrary function \cite{hybridmetricpalatini}, or further to $f(R,\mathcal{R})$ \cite{generalizedHMP}. Extensive studies have been done in the literature exploring cosmological \cite{cosHMP}, balck holes \cite{BlackHMP} and theoretical \cite{theoHMP} aspects of the theory. In summary, these types of theories can be seen as a multi scalar-tensor generalization of the Einstein general relativity \cite{hybridmetricpalatini}.

In this paper, we will generalize the hybrid metric-Palatini idea to contain extra higher order terms in the Palatini sector. As a result we will consider a Lagrangian of the form $R+f(\mathcal{R},\mathcal{R}_{\mu\nu} \mathcal{R}^{\mu\nu})$, where the Palatini Ricci tensor is explicitly included in the theory. The main difference between this model and the original hybrid metric-Palatini model is that here, there is no scalar-tensor alternative. In fact, because of the existence of $\mathcal{R}_{\mu\nu}$ in the action, the secondary metric is not conformally related to the original metric, make it impossible to find a scalar-tensor alternative for the model. In this sense, the present model has a reacher structure compared to the original hybrid metric-Palatini model. We will see the result of this extension in the cosmological predictions of the universe. The present model can explain the recent observational data very good; at the lower orders it is very similar to the $\Lambda$CDM model. However, in order to explore more the differences between this model and the $\Lambda$CDM model, we will perform higher order analysis, including the statefinder and the $Om$ diagnostics. As a result the model is different from $\Lambda$CDM model but can explain the observations as well.

The paper is organized as follows: In the nest section we will present the model and try to simplify it on top of a maximally symmetric space-times. In section \ref{cos} we will obtain the cosmological equations and perform the likelihood analysis to obtain the cosmological and model parameters in confrontation with the observational data on the Hubble parameter from the cosmic chronometers and Pantheon+ datasets. We will then explore the cosmological implications of the model in details and perform the higher derivative diagnostics to compare the model with $\Lambda$CDM model. In the last section, we will conclude the paper.
\section{The model}
Let us consider a Lagrangian density of the form
\begin{align}\label{action}
S=\int d^4x\sqrt{-g}\,\Big[\kappa^2 \big(R+f(\mathcal{R},\mathcal{R}_{\mu\nu} \mathcal{R}^{\mu\nu})\big)+L _m\Big],
\end{align}
where $L _m$ is the matter Lagrangian. The first term is the metric Ricci scalar and the function $f$ is an arbitrary function of the Palatini Ricci scalar $\mc{R}$ and Ricci tensor $\mc{R}_{\mu\nu}$. In this sense, the first Ricci scalar in \eqref{action}, is constructed only by the metric field $g_{\mu\nu}$; the connections being the Christoffel symbol. However, the arguments in the function $f$ are constructed by the metric field $g_{\mu\nu}$ and an independent affine connection $\hat\Gamma^\mu_{~\nu\alpha}$. 

By variation of the action \eqref{action} with respect to the metric field $g_{\mu\nu}$ and denoting $Q=\mathcal{R}_{\mu\nu} \mathcal{R}^{\mu\nu}$, one could obtain the metric field equation as
\begin{align}\label{field1}
G_{\mu\nu} -\frac{1}{2} g_{\mu\nu} f+ f_{\mathcal{R}}\mathcal{R}_{\mu\nu}+2f_{Q} \mathcal{R}_{\mu\alpha} \mathcal{R}^{\alpha}_{\nu}=\f{1}{2\kappa^2} T_{\mu\nu} ,
\end{align}
where $T_{\mu\nu}$ is the energy-momentum tensor defined as
\begin{align}\label{emtensor}
	T_{\mu\nu}=-\frac{2}{\sqrt{-g}}\frac{\delta(\sqrt{-g}L_m )}{\delta g^{\mu\nu}}.
\end{align}
Also the variation of the action \eqref{action} with respect to the affine connection $\hat\Gamma^\mu_{~\nu\alpha}$ gives the connection field equation
\begin{align}\label{field2}
\hat\nabla_{\beta}\Big[ \sqrt{-g}(  f_{\mathcal{R}}g^{\mu\nu}+2f_{Q} \mathcal{R}^{\mu\nu})\Big]=0.
\end{align}
where $f_i$ denotes derivation of the function $f$ with respect to the quantity $i$. It should be noted that $\hat\nabla_\beta$ is the covariant derivative with respect to the independent connection $\hat\Gamma^\mu_{~\nu\alpha}$. 

In the hybrid metric-Palatini gravity, we have two different covariant derivatives; one is constructed from the Levi-Civita connection with the property $\nabla_\alpha g_{\mu\nu}=0$, and the other is constructed from the affine connection $\hat\Gamma^\mu_{~\nu\alpha}$ which is generally metric incompatible, e.g. $\hat\nabla_\alpha g_{\mu\nu}\neq0$. 

In Palatini-based models, it is customary to define a new metric $h_{\mu\nu}$ using the connection equation, with the property $\hat\nabla_\alpha h_{\mu\nu}=0$ through which we can obtain the Palatini curvature tensors $\mathcal{R}_{\mu\nu}$ and $\mathcal{R}$ straightforwardly.
One can define such a metric tensor $h_{\mu\nu}$ using the structure of the connection equation \eqref{field2} as
\begin{align}\label{hmetric}
\sqrt{-h}h^{\mu\nu}&=\sqrt{-g}(  f_{\mathcal{R}}g^{\mu\nu}+2f_{Q} \mathcal{R}^{\mu\nu})
\nonumber\\&=\sqrt{-g}\hat\Sigma^{\mu\nu},
\end{align}
with the property $\hat\nabla_\beta(\sqrt{-h}h^{\mu\nu})=0$. In the above expression, we have defined a dimensionless tensor field $\hat{\Sigma}^\mu_\nu$ as
\begin{align}
\hat{\Sigma}^\mu_\nu&=2f_{Q}\mc{R}^\mu_\nu+f_{\mathcal{R}}\delta^\mu_\nu.
\end{align}
As one can see from equation \eqref{hmetric}, the metric $h_{\mu\nu}$ proportional to the Ricci tensor $\mathcal{R}_{\mu\nu}$. This implies that the present model takes into account all the components of the Palatini Ricci tensor, in contrary to other hybrid metric-Palatini models, such as $f(R,\mathcal{R})$, where the metric $h_{\mu\nu}$ is proportional to $g_{\mu\nu}$ and reflects only the properties of the Ricci scalar $\mathcal{R}$. 

Noting that the covariant derivative $\hat\nabla_\beta$ is $h_{\mu\nu}$-metric compatible, one can obtain the connection coefficients $\hat\Gamma^\alpha_{~\beta\gamma}$ as
\begin{align}\label{conn1}
\hat\Gamma^\alpha_{~\beta\gamma}=\f12h^{\alpha\delta}\Big(\partial_\beta h_{\delta\gamma}+\partial_\gamma h_{\beta\delta}-\partial_\delta h_{\beta\gamma}\Big).
\end{align}
Generally, by solving equation \eqref{hmetric} and obtaining the explicit relation for the metric $h_{\mu\nu}$ in terms of $g_{\mu\nu}$, one could obtain the connection coefficients in terms of the metric tensor $g_{\mu\nu}$ which enables us to express the Palatini Ricci tensor $\mathcal{R}_{\mu\nu}$ and Ricci scalar $\mathcal{R}$ in term of the metric $g_{\mu\nu}$. In this sense, equation \eqref{field1} governs only the metric field $g_{\mu\nu}$.  

In the case of $f=f(\mathcal{R})$, we have $f_Q=0$ and the model can be rewritten in the form of a scalar-tensor theory \cite{hybridmetricpalatini}. For a general case with $f_Q\neq0$, equation \eqref{hmetric} can not be solved analytically for $h_{\mu\nu}$. However, for maximally symmetric space-times such as the FRW universe, this could be done as we will address in the following.

Let us first assume that the matter content of the universe can be described by a perfect fluid with Lagrangian density $L _m=-\rho$ and energy-momentum tensor
\begin{align}\label{EM}
	T_{\mu\nu}=(p+\rho )u_\mu u_\nu+pg_{\mu\nu},
\end{align}
where $\rho$ and $p$ are energy-density and thermodynamics pressure respectively.
Now, define a tensor $\tau_{\mu\nu}$ as
\begin{align}
\tau_{\mu\nu}\equiv T_{\mu\nu}-2\kappa^2 G_{\mu\nu}.
\end{align}
For maximally symmetric space-times, the tensor $\tau_{\mu\nu}$ has a structure similar to \eqref{EM} as
\begin{align}\label{tau}
\tau_{\mu\nu}=(\rho_\tau+p_\tau)u_\mu u_\nu+p_\tau g_{\mu\nu},
\end{align}
where $\rho_\tau$ and $p_\tau$ are the extended energy density and pressure which depend on the matter content and also on the metric tensor $g_{\mu\nu}$.

Using the decomposition \eqref{tau}, one can rewrite the metric equation \eqref{field1} as
\begin{align}\label{rel1}
\left(2f_{Q}\hat{\mc{R}}+\frac{f_{\mathcal{R}}}{2}I\right)^2= \alpha _{1}I+\alpha _{2}U,
\end{align}
where $\hat{\mc{R}}$ is the matrix form of the Palatini Ricci tensor $\mc{R}^\mu_\nu$, $I$ is the unit $4\times4$ matrix and we have defined the $4\times4$ matrix $U$ as
$U^\mu_\nu=u^\mu u_\nu$.
The coefficients $\alpha_1$ and $\alpha_2$ can be obtained as
\begin{align}
\alpha_1&=f_Q\left(f+\frac{p_\tau}{2\kappa^2}\right) +\frac14f^{2}_{\mathcal{R}} ,\nonumber\\
\alpha_2&=\frac{1}{2\kappa^2}f_{Q}\big(\rho_\tau+p_\tau\big).
\end{align}
The quantities $\alpha_1$ and $\alpha_2$ are both dimensionless.

As observed thus far, all second-rank tensors in the model are expressed in terms of \( I \) and \( U \). Consequently, we assume that the square root of the left-hand side of equation \eqref{rel1} can also be expanded as  
\begin{align}\label{rel2}
	2f_{Q}\hat{\mc{R}}+\frac{f_{\mathcal{R}}}{2}I = \bar\alpha _{1}I+\beta_{2}U.
\end{align}
By squaring equation \eqref{rel2} and comparing it with \eqref{rel1}, one obtains  
\begin{align}
	\alpha_1 = \bar\alpha_1^2, \qquad \beta_2 = \sqrt{\alpha_{1}} - \sqrt{\alpha_{1} - \alpha_{2}}.
\end{align}
As a result, the Palatini Ricci tensor can be written as  
\begin{align}\label{R1}
	\mc{R}^\mu_\nu = \frac{1}{2f_Q} \left( \beta_{1} \delta^\mu_\nu + \beta_{2} u^\mu u_\nu \right),
\end{align}
where we have defined
\begin{align}
\beta_{1}&=\sqrt{\alpha_{1}}-\frac{f_{\mathcal{R}}}{2}.
\end{align}
The above expression imposes that
\begin{align}\label{con1}
\alpha_1>0,\qquad\alpha_1>\alpha_2,
\end{align}
which can be expanded as
\begin{align}
f_R^2&>max\big\{-4f_Q(2\kappa^2f-\rho_\tau),-4f_Q(2\kappa^2f+ p_\tau)\big\}. \nonumber\\
\end{align}
Using equation \eqref{R1}, one can obtain the matrix $\hat\Sigma$ as
\begin{align}\label{S1}
\hat{\Sigma}=\beta_3I+\beta_{2}U,
\end{align}
where we have defined
\begin{align}
\beta_{3}=\beta_{1}+f_{\mathcal{R}}=\sqrt{\alpha_{1}}+\frac{f_{\mathcal{R}}}{2}.
\end{align}
All $\beta_i$'s are dimensionless quantities.
The inverse of the matrix $\hat\Sigma$ and its determinant can be obtained from \eqref{S1} as
\begin{align}
\hat{\Sigma}^{-1}=\frac{1}{\beta_{3}}I+\frac{\beta_{2}}{\beta_{3}(\beta_{2}-\beta_{3})}U,
\end{align}
and
\begin{align}
\det\hat\Sigma=\beta_{3}^{3}(\beta_{3}-\beta_{2}),
\end{align}
Now, here is the right place to obtain the form of the metric $h_{\mu\nu}$. Notice that equation \eqref{hmetric} can be rewritten in a matrix form as
\begin{align}\label{hmetric2}
	\hat{h}^{-1}=\frac{\sqrt{-g}}{\sqrt{-h}}\,\hat\Sigma\, \hat{g}^{-1}.
\end{align}
Taking the determinant of the above relation gives
\begin{align}
	\textmd{det} \hat\Sigma=\frac{\textmd{det}h}{\textmd{det}g},
\end{align}
Substituting back to \eqref{hmetric2} gives the matrix form of the metric $h_{\mu\nu}$ as
\begin{align}
	\hat{h}=\sqrt{det\hat\Sigma}\,\hat{g}\,\hat{\Sigma}^{-1}.
\end{align}
Combining the above relations, one can obtain the metric $h_{\mu\nu}$ as
\begin{align}\label{metrich}
	h_{\mu\nu}=\epsilon\sqrt{\beta_{3}(\beta_{3}-\beta_{2})} g_{\mu\nu}-\frac{\beta_2|\beta_3|}{\sqrt{\beta_{3}(\beta_{3}-\beta_{2})}}u_\mu u_\nu,
\end{align}
where $\epsilon=sign(\beta_3)$ is the sign of $\beta_3$ and we have a constraint
\begin{align}
	\beta_3(\beta_3-\beta_2)>0.
\end{align}
It should be noted that in the expression \eqref{metrich} for the metric \( h_{\mu\nu} \), the 4-velocity of the matter field appears. This arises from our specific decomposition of tensors in a form of a perfect fluid. Since we use the metric field equation \eqref{field1} to relate the Palatini metric \( h_{\mu\nu} \) to \( g_{\mu\nu} \), all dynamical fields depend on both the matter sector and the geometry. We emphasize again that the current procedure is valid only in the FRW universe, where all tensors can be expressed in the form of a perfect fluid.

Since the signature of metric $h_{\mu\nu}$ should be the same as $g_{\mu\nu}$ we demand $\epsilon=1$ which implies that
\begin{align}\label{cons2}
\beta_3>0,\qquad \beta_3>\beta_2.
\end{align}
From the second constraint, one obtains
\begin{align}
	f_R>0.
\end{align}
As a result, the metric $h_{\mu\nu}$ reduces to
\begin{align}
h_{\mu\nu}=\gamma_2g_{\mu\nu}-\gamma_1u_{\mu}u_{\nu}.
\end{align}
where we have defined
\begin{align}
\gamma_{1}=\beta_{2}\sqrt{\frac{\beta_{3}}{\beta_{3}-\beta_{2}}},\quad
\gamma_{2}=\sqrt{\beta_{3}(\beta_{3}-\beta_{2})}.
\end{align}
The above metric can then be used to compute the connection coefficients from \eqref{conn1}, and the Palatini Ricci tensor $\mc{R}_{\mu\nu}$ and Ricci scalar $\mc{R}$. These tensors will then be used in equation \eqref{field1} to obtain gravitational field equations. In the next section, we will perform the above procedure for an FRW universe filled with dust.

\section{Cosmology}\label{cos}
Let us apply the aforementioned procedure to an isotropic and homogeneous universe described by the FRW line element of the form
\begin{align}
ds^{2}=-dt^{2}+a^{2}d\vec{x}^{2},
\end{align}
where $a=a(t)$ is the scalar factor. The velocity 4-vector of the matter fluid can be written in FRW space-times as
\begin{align}
u^{\mu}=\left(1,0,0,0\right).
\end{align}
In the case of FRW universe, the Einstein tensor can be decomposed in a perfect fluid form as
\begin{align}
G^{\mu}_{\nu}=-\left(3H^2+2\dot{H}\right)\delta ^{\mu}_{\nu}-2\dot{H} u^{\mu}u_{\nu}.
\end{align}
 Using the above expression, one can write the tensor $\tau^{\mu}_{~\nu}$ as 
\begin{align}\label{tau1}
\tau ^{\mu}_{~\nu}=\left(\rho_{\tau} + p_{\tau} \right)u^{\mu}u_{\nu}+p_{\tau}\delta ^{\mu}_{\nu},
\end{align}
where
\begin{align}
\rho_\tau&=\rho - 6\kappa ^{2}H^2,\\
p_\tau&=p +4\kappa ^{2}\dot{H} +6\kappa ^{2}H^2.
\end{align}
The line element of the h-metric can be written in the FRW space-time as
\begin{align}\label{hmet}
	ds^{2}_{h}=-(\gamma_{1}+\gamma _{2})dt^{2}+\gamma_{2}a^{2}d\vec{x}^{2}.
\end{align}
One can see that the metric $h_{\mu\nu}$ has also an FRW form 
\begin{align}
	ds_h^{2}=-N^2dt^{2}+\tilde{a}^{2}d\vec{x}^{2},
\end{align}
with
\begin{align}
	N=\sqrt{\gamma_{1}+\gamma_{2}},\quad
	\tilde{a}=\sqrt{\gamma_2}a.
\end{align}
It should be noted that from the constraint equations \eqref{cons2} the above functions are well defined.

The related Palatini quantities can be computed from the metric \eqref{hmet} as
\begin{align}
	\mathcal{R}^\mu_\nu&=\left(3K^2+\frac{\dot{K}}{N}\right)\delta^\mu_\nu-\frac{2\dot{K}}{N}u^\mu u_\nu,\nonumber\\
	\mathcal{R}&=6\left(2K^2+\frac{\dot{K}}{N}\right),
\end{align}
and
\begin{align}
	Q\equiv\mathcal{R}_{\mu\nu}\mathcal{R}^{\mu\nu}=12\left(3K^4+\frac{3K^2}{N}\dot{K}+\frac{\dot{K}^2}{N^2}\right),
\end{align}
where we have defined the h-Hubble parameter $K$ as
\begin{align}
	K=\frac{\dot{\tilde{a}}}{\tilde{a}N}.
\end{align}
In terms of the Hubble parameter, we have
\begin{align}
	K = \frac{1}{N}\left(H+\frac{\dot{\gamma_2}}{2\gamma_2}\right).
\end{align}
As a result, from the metric equation \eqref{field1}, one can obtain the Friedmann and Raychaudhuri equations as
\begin{align}\label{fe1}
3H^2&+\f12f-3K^2(f_\mathcal{R}+6f_QK^2)\nonumber\\&-\frac{3\dot{K}}{N^2}\left[(f_\mathcal{R}+12f_QK^2)N+6f_Q\dot{K}\right]=\frac{1}{2\kappa^2}\rho ,
\end{align}
and
\begin{align}\label{fe2}
	2\dot{H}+\frac{2\dot{K}}{N^2}\left[(f_\mathcal{R}+12f_QK^2)N+8f_Q\dot{K}\right]=\frac{1}{2\kappa^2}(\rho+p).
\end{align}
The hybrid metric-Palatini model in this paper is meant to be be an alternative to dark energy. As a result, let us consider more on the behavior of the dark fluid described by the hybrid metric-Palatini model. Let us define the effective energy-density and pressure by rewriting the Friedmann and Raychaudhuri equations as
\begin{align}\label{Fr1}
	3H^2=\frac{1}{2\kappa^2} (\rho+\rho_{eff}),
\end{align}
and
\begin{align}\label{Fr2}
	2\dot{H}+3H^2=-\frac{1}{2\kappa^2}(p+p_{eff}),
\end{align}
For the hybric metric-Palatini model, one can obtain
\begin{align}\label{reff}
	\frac{1}{\kappa^2}\rho_{eff} = &-f+6K^2(f_{\mathcal{R}}+6f_QK^2)\nonumber\\&+\frac{6\dot{K}}{N^2}\left[(f_{\mathcal{R}}+12f_QK^2)N+6f_Q\dot{K}\right],
\end{align}
and
\begin{align}\label{peff}
	\frac{1}{\kappa^2}p_{eff} = &f-6K^2(f_{\mathcal{R}}+6f_QK^2)\nonumber\\&-\frac{2\dot{K}}{N^2}\left[(f_{\mathcal{R}}+12f_QK^2)N+2f_Q\dot{K}\right].
\end{align}
Let us now discuss the conservation of the energy-momentum tensor. In practice, matter conservation can be derived via Noether’s theorem from the diffeomorphism invariance of the action \cite{noether}; see \cite{noetherpalatini} for such an application in theories involving the Palatini variation. In our case, considering the action \eqref{action}, one can see that there are three diffeomorphism-invariant terms, each of which should be independently invariant under a diffeomorphism transformation of the form:
\begin{align}\label{diff}
	x^\mu\rightarrow x^{\prime\mu}=x^\mu+\xi^\mu(x).
\end{align}
Let us focus on the last term in \eqref{action}. The variation of this term could be obtained as
\begin{align}\label{eqcons1}
	\delta_\xi S = \int d^x\left(\frac{\delta (\sqrt{-g}L_m)}{\delta g^{\mu\nu}}\delta g^{\mu\nu}+\frac{\delta (\sqrt{-g}L_m)}{\delta \Phi^a}\delta\Phi^a\right),
\end{align}
where we have assumed that the matter Lagrangian $L_m$ depends on the metric tensor and some matter fields denoted by $\Phi^a$.
Since matter Lagrangian $L_m$ only appears in this term, one can deduce that the quantity
$$\frac{\delta (\sqrt{-g}L_m)}{\delta \Phi^a},$$
represents the equation of motion for the matter sector and thus vanishes when the matter field equations are satisfied. This is a crucial observation because, in theories with non-minimal matter-geometry couplings, $L_m$ appears in other terms of the action, and the above expression does not generally vanish.
Noting that the variation of the metric tensor under diffeomorphism \eqref{diff} is given by
\begin{align}
	\delta g_{\mu\nu} \equiv \mathcal{L}_\xi g_{\mu\nu} =2 \nabla_{(\mu}\xi_{\nu)},
\end{align}
one can write the variation \eqref{eqcons1} as
\begin{align}
		\delta_\xi S = -2\int d^x\nabla^\mu\left(\frac{\delta (\sqrt{-g}L_m)}{\delta g^{\mu\nu}}\right)\xi^\nu.
\end{align}
Now, using the definition of the energy-momentum tensor \eqref{emtensor} and requiring that the variation vanishes due to diffeomorphism invariance, one obtains
$$\nabla_\mu T^{\mu\nu}=0.$$
On top of FRW universe, the energy-momentum conservation equation reduces to
\begin{align}\label{cons}
	\dot{\rho}+3H(\rho+p)=0.
\end{align}
In the above equations, independent variables are the scale factor $a$ and the energy density $\rho$. 
\begin{figure*}
	\includegraphics[scale=0.4]{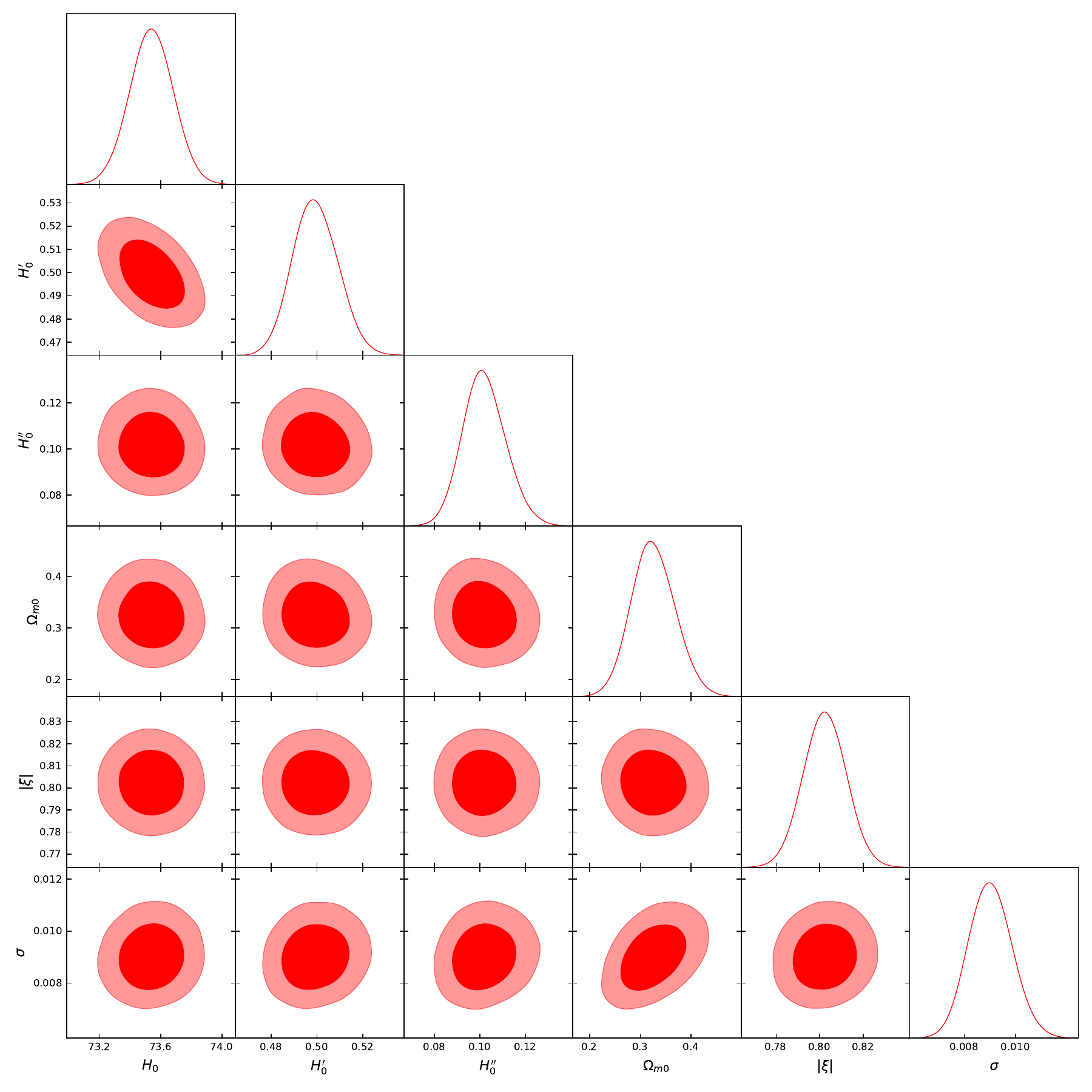}
	\caption{The corner plot for the values of the parameters $H_0$,$H^\prime_0$, $h^{\prime\prime}_0$, $\Omega_{m0}$, $\xi$ and $\tilde{\sigma}$ with their $1\sigma$ and $2\sigma$ confidence levels for the hybrid metric-Palatini $\sigma\mathcal{R}+\xi\mathcal{R}_{\mu\nu}\mathcal{R}^{\mu\nu}$ gravity model. \label{cornerplot}}
\end{figure*}
From the conservation equation \eqref{cons}, the behavior of the energy density $\rho$ is related to the scale factor $a$ as
\begin{align}
	\rho = \frac{\rho_0}{a^{3(1+\omega)}},
\end{align}
where $\omega=p/\rho$ is the equation of state parameter and $\rho_0$ is the energy density at present time.

The $\mathcal{R}$ and $Q$ scalars which appears in the Friedmann and Raychaudhuri equations should be obtained as a function of $a$ and $\rho$ from the following set of equations
\begin{align}\label{ex1}
	\mathcal{R}&=\frac{1}{2f_Q}\left(3A+B-2f_\mathcal{R}\right),\nonumber\\
	Q&=\frac{1}{4f_Q^2}\left(3A^2+B^2-(3A+B)f_\mathcal{R}+f_\mathcal{R}^2\right),
\end{align}
which are obtained from equation \eqref{R1}. Here, we have denoted
\begin{align}\label{ex2}
	A\equiv\sqrt{\alpha_1},\qquad B\equiv\sqrt{\alpha_1-\alpha_2}.
\end{align}
\begin{figure*}
	\includegraphics[scale=0.5]{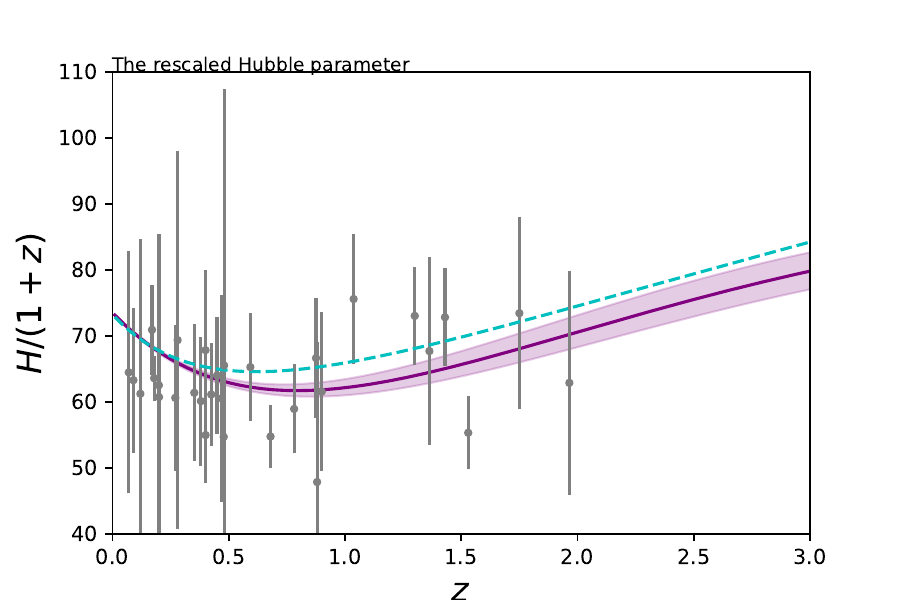}\includegraphics[scale=0.5]{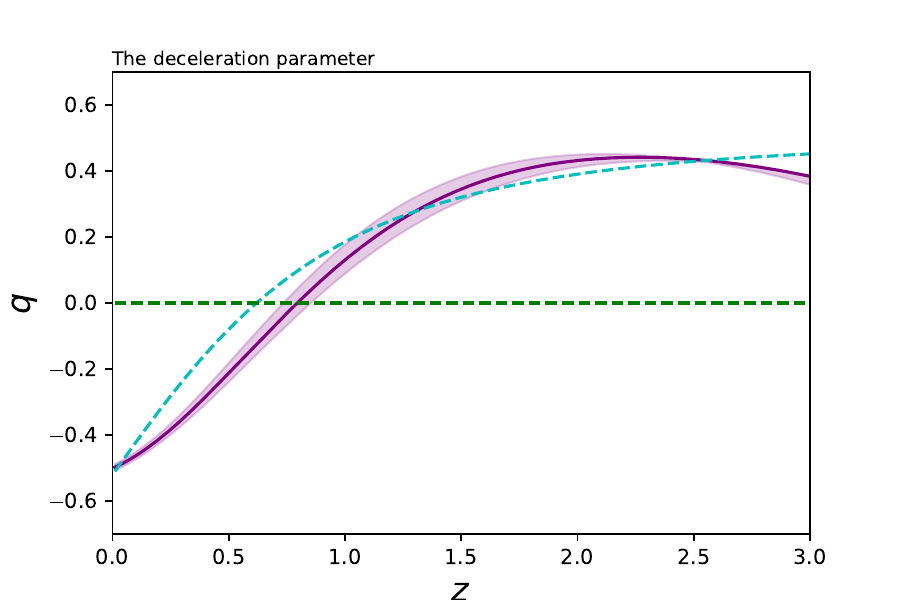}
	\caption{\label{fighubq} The behavior of the rescaled Hubble parameter $H/(1+z)$ (left panel) and of the deceleration parameter $q$ (right panel) as a function of the redshift for the metric-Palatini $\sigma\mathcal{R}+\xi\mathcal{R}_{\mu\nu}\mathcal{R}^{\mu\nu}$ gravity model for the best fit values of the parameters as given by table \eqref{bestfit}. The shaded area denotes the $1\sigma$ error. Dashed lines represent $\Lambda$CDM model.}
\end{figure*}
Let us now assume that the function $f$ has a linear structure on its arguments as
\begin{align}
f(\mathcal{R},\mathcal{R}_{\mu\nu}\mathcal{R}^{\mu\nu})=\xi\mathcal{R}+\sigma \mathcal{R}_{\mu\nu}\mathcal{R}^{\mu\nu},
\end{align}
where $\xi$ and $\sigma$ are some arbitrary constants. 
In this case, one can obtain
\begin{align}
	\alpha_1=\sigma\left(3H^2+2\dot{H}+\frac{1}{2\kappa^2}p\right)+\sigma(\xi\mathcal{R}+\sigma Q)+\f14\xi^2,
\end{align}
and
\begin{align}
	\alpha_1-\alpha_2=\sigma\left(3H^2-\frac{1}{2\kappa^2}\rho\right)+\sigma(\xi\mathcal{R}+\sigma Q)+\f14\xi^2.
\end{align}
Now, assuming that the universe is filled with dust with equation of state $p=0$ and using the expressions \eqref{ex1} and \eqref{ex2}, one can obtain $A$, $B$, $\mathcal{R}$ and $Q$ as functions of $H$ and $\rho$
\begin{align}
	A=\frac{1}{16\kappa^2\xi}\left[3\sigma\rho+12\kappa^2(\xi^2-6\sigma^2H^2-3\sigma\dot{H})-C\right],
\end{align}
\begin{align}
	B=-\frac{1}{16\kappa^2\xi}\left[\sigma\rho+4\kappa^2(\xi^2-6\sigma^2H^2-3\sigma\dot{H})-C\right],
\end{align}
\begin{align}
	\mathcal{R}=-\f{1}{\xi}\left(6H^2+3\dot{H}-\frac{1}{4\kappa^2}\rho\right),
\end{align}
and
\begin{align}
	Q=\f{1}{8\sigma^2}\Bigg(12\sigma H^2&-\frac{3}{4\kappa^2\xi}C(\xi+\sigma\mathcal{R})-\frac{2\sigma}{\kappa^2}\rho\nonumber\\&+(\xi+\sigma\mathcal{R})(3\xi+5\sigma\mathcal{R})\Bigg),
\end{align}
where we have defined
\begin{align}
	C=&\Big(16\kappa^4(\xi^2-6\sigma H^2)^2-8\kappa^2\sigma\rho(\xi^2+6\sigma H^2)\nonumber\\&+8\kappa^4\sigma \dot{H}(72\sigma H^2-20\xi^2)+24\kappa^2\sigma^2\dot{H}(6\kappa^2\dot{H}-\rho)\Big)^{\frac{1}{2}}.
\end{align}
We are now ready to solve the cosmological field equations \eqref{fe1} and \eqref{fe2} together with the conservation equation \eqref{cons}. 
Let us define the following set of dimensionless quantities
\begin{align}
	\tau &=H_0t, \quad H=H_0 h,\quad \tilde{\rho}=\f{1}{6\kappa^2H_0^2}\rho,\quad
	\tilde{\sigma}=\sigma H_0^{-2}.
\end{align}
and transform to the redshift coordinates defined as
\begin{align}
	\f{1}{a}=1+z,
\end{align}
which implies
\begin{align}
	 \frac{d}{d\tau} = \frac{d}{dz}\frac{dz}{d\tau}=-(1+z)h(z)\frac{d}{dz}.
\end{align}
\begin{figure*}
	\includegraphics[scale=0.5]{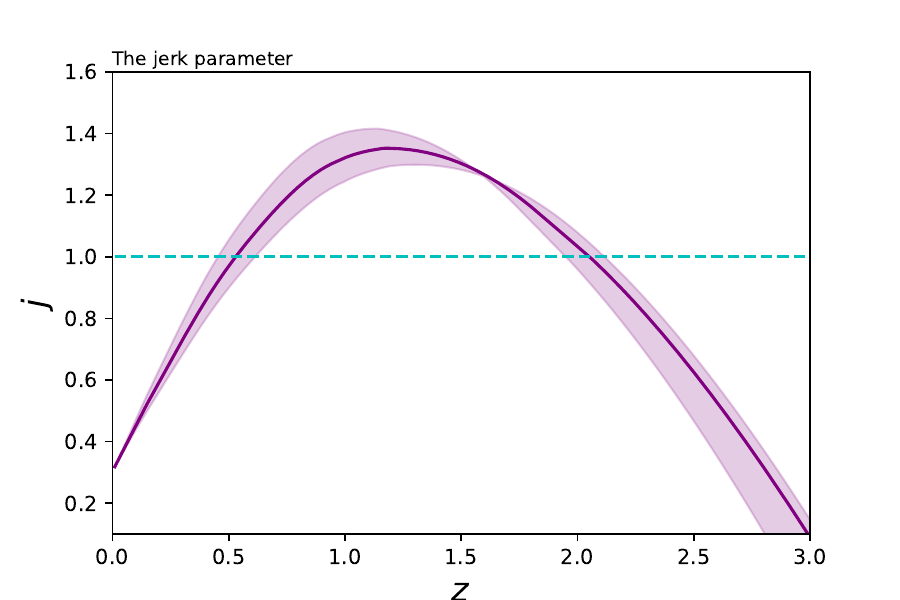}\includegraphics[scale=0.5]{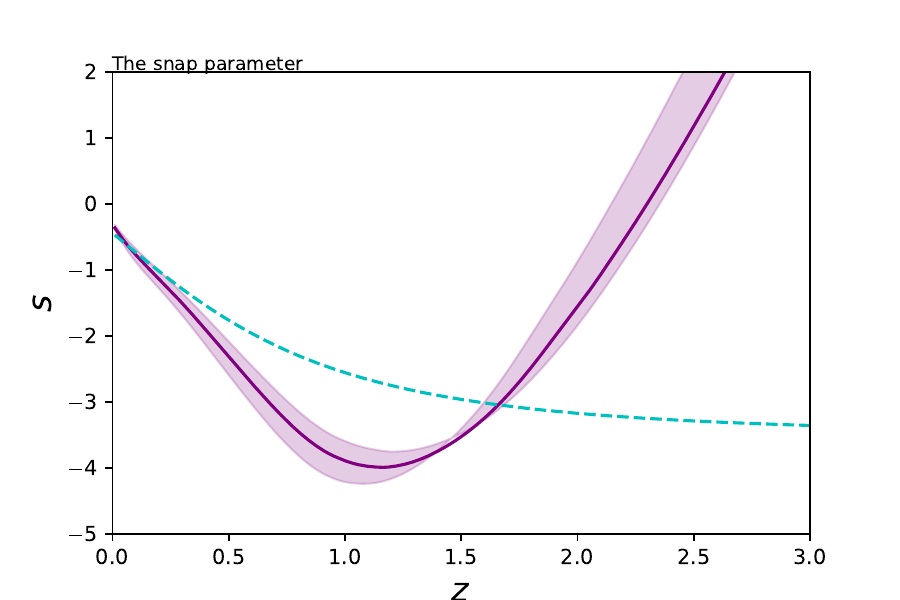}
	\caption{\label{figjands} The behavior of the jerk $j$ (left panel) and of the snap parameter $s$ (right panel) as a function of the redshift for the metric-Palatini $\sigma\mathcal{R}+\xi\mathcal{R}_{\mu\nu}\mathcal{R}^{\mu\nu}$ gravity model for the best fit values of the parameters as given by table \eqref{bestfit}. The shaded area denotes the $1\sigma$ error. Dashed lines represent $\Lambda$CDM model.}
\end{figure*}
\begin{figure*}
	\includegraphics[scale=0.47]{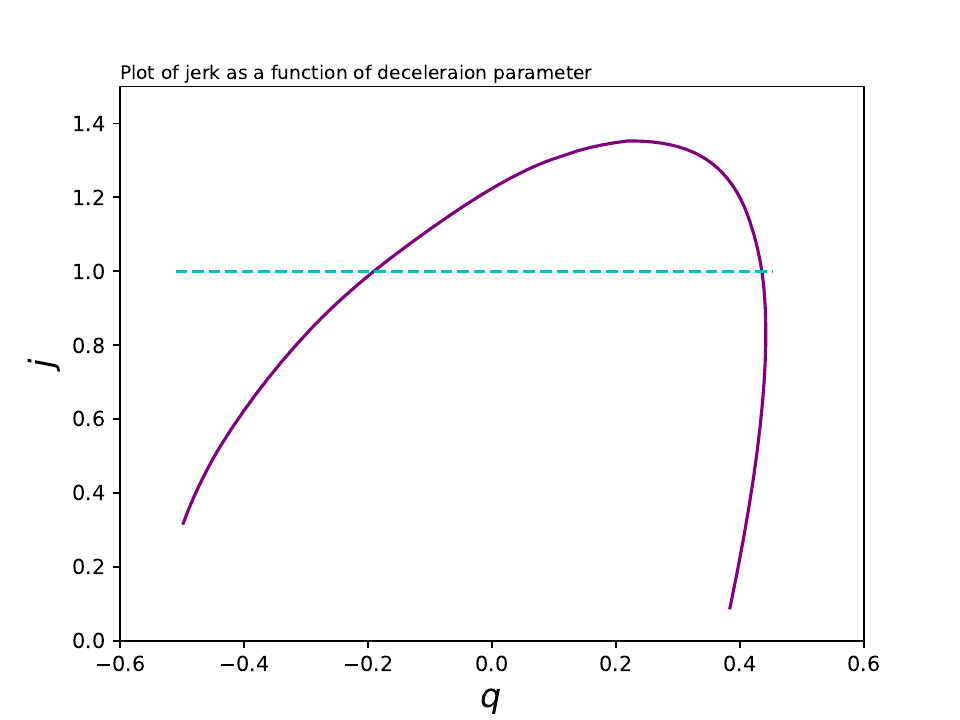}\includegraphics[scale=0.47]{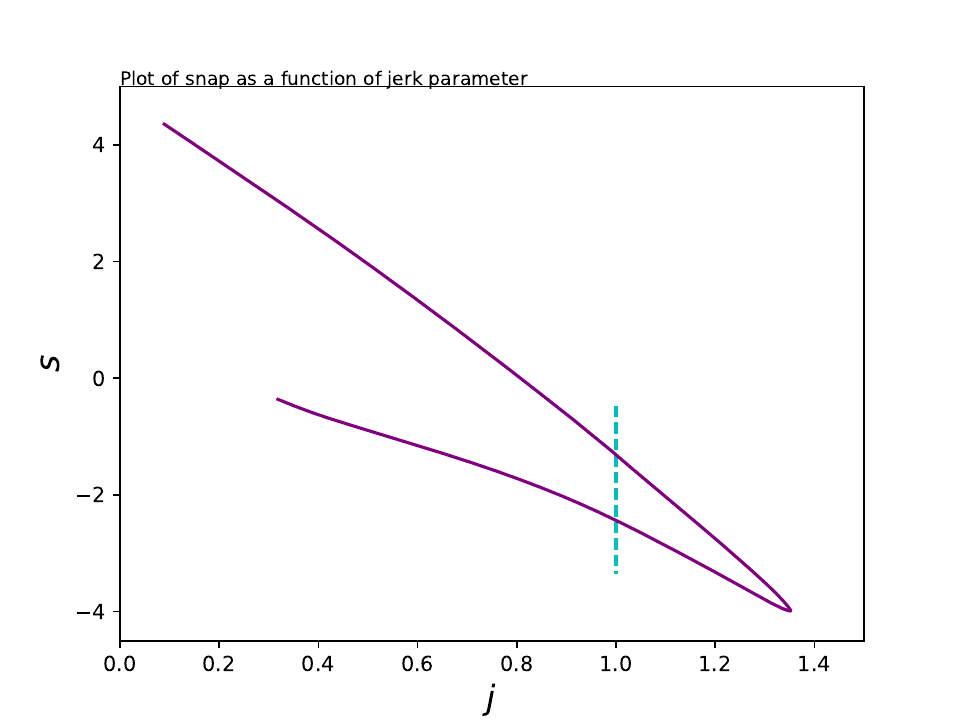}
	\caption{\label{figversus} The variation of the jerk parameter $j$ as a function of the deceleration parameter $q$, $j=j(q)$ (left panel), and of the snap parameter as a function of the jerk parameter, $s=s(j)$ (right panel) for the metric-Palatini $\sigma\mathcal{R}+\xi\mathcal{R}_{\mu\nu}\mathcal{R}^{\mu\nu}$ gravity model for the best fit values of the parameters as given by table \eqref{bestfit}. Dashed lines represent $\Lambda$CDM model.}
\end{figure*}
We should note that the model has only one independent equation which governs the behavior of the Hubble parameter. We will not write the detailed equation here for brevity. But for small values of the parameter $\tilde\sigma$, the evolution equation of the Hubble parameter  can be expanded as
\begin{align}
(2hh^\prime&+3\Omega_{m 0}(1+z)^2)\xi^2\nonumber\\&+2(1+z)^2\tilde{\sigma}h^2(9\Omega_{m 0}-4h^\prime h^{\prime\prime})\nonumber\\&-2(1+z)\tilde{\sigma}h^3(4h^{\prime\prime}+(1+z)h^{\prime\prime\prime})\nonumber\\
&+4\tilde{\sigma}hh^\prime(11(1+z)hh^\prime-43h^2)\nonumber\\
&+2(1+z)^2\tilde{\sigma}hh^\prime(6\Omega_{m0}(1+z)-h^{\prime2})=0.
\end{align}

The deceleration parameter for the cosmic expansion of the FLRW metric gives
\begin{align}
	q = -1 +(1+z)\frac{h^\prime}{h}.
\end{align}
\subsection{Statistical analysis}
In order to obtain the evolution of the Hubble parameter, the field equation must be integrated with initial conditions $h(0)=1$, $h^\prime(0)=h^\prime_0=H^\prime_0$ and $h^{\prime\prime}(0)=h^{\prime\prime}_0$. We will obtain the best fit values of the parameters $H_0$, $H^\prime_0$, $H^{\prime\prime}_0$, $\Omega_{m0}$, $\xi$ and $\tilde{\sigma}$, by performing the Likelihood analysis using 31 data points related to the cosmic chronometers (CC) \cite{CCdata} together with the Pantheon+ measurements with SH0ES calibration \cite{PANdata}. We will assume that the CC data are independent but the Pantheon+SH0ES data are correlated \cite{PANdata}.

The likelihood function can then be defined as
\begin{align}
	L=L_0e^{-\chi^2/2},
\end{align}
where $L_0$ is the normalization constant. The loss functions $\chi^2$ for the cosmic chronometer and Pantheon+ data points are defined as
\begin{align}
	\chi_{CC}^2=\sum_i\left(\frac{{H}_{\text{obs},i}-{H}_{\text{th},i}}{\sigma_i}\right)^2,
\end{align}
and 
\begin{equation}
	\chi^2_{\text{Pantheon+SH0ES}} = \left[\vec{\mu}_{\text{obs}} - \vec{\mu}_{\text{th}}\right]^T 
	C^{-1} 
	\left[\vec{\mu}_{\text{obs}} - \vec{\mu}_{\text{th}}\right]
\end{equation}
Here $i$ counts data points, $``obs"$ are the observational values of the Hubble parameter, $``th"$ are the theoretical values obtained from the model, and $\sigma_i$ are the errors associated with the $i$th data obtained from observations. Also, $C$ is the covariance matrix associated with Pantheon+ datapoints \cite{PANdata}.

By maximizing the likelihood function, the best fit values of the parameters $H_0$,$H^\prime_0$, $\Omega_{m0}$, $\xi$ and $\tilde{\sigma}$ at $1\sigma$ confidence level, can be obtained as
\begin{table}[h!]
	\centering
	\begin{tabular}{|c|c|}
		\hline
		Parameter & best fit value with 1$\sigma$ error  \\\hline\hline
		$H_0$ & $73.538^{+0.1418}_{-0.1417}$ \\\hline
		$H^\prime_0$ & $0.499^{+0.0102}_{-0.0095}$ \\\hline
		$h^{\prime\prime}_0$ & $0.102^{+0.0098}_{-0.0089}$ \\\hline
		$\Omega_{m0}$ & $0.324^{+0.0446}_{-0.0413}$ \\\hline
		$\xi$ & $-0.802^{+0.0098}_{-0.0098}$ \\\hline
		$\tilde\sigma$ & $0.009^{+0.0009}_{-0.0008}$ \\\hline
	\end{tabular}
	\caption{The best fit values together with their 1$\sigma$ errors for the hybrid metric-Palatini gravity.}\label{bestfit}
\end{table}
\begin{figure}[h!]
	\includegraphics[scale=0.5]{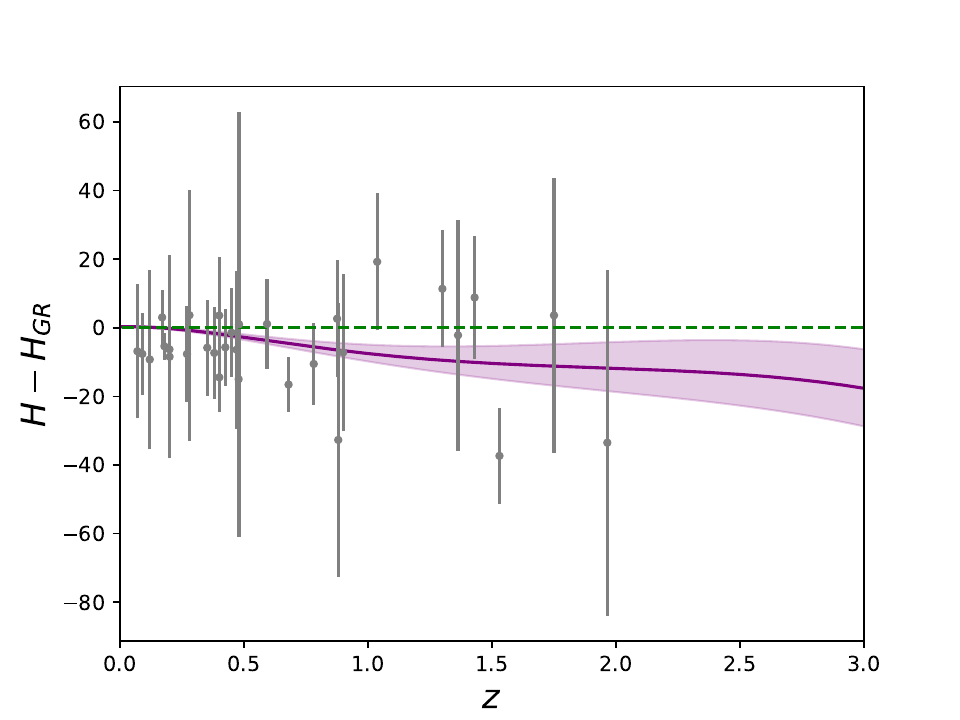}
	\caption{\label{hdiff} The difference between the metric-Palatini $\sigma\mathcal{R}+\xi\mathcal{R}_{\mu\nu}\mathcal{R}^{\mu\nu}$ gravity model and its $\Lambda$CDM counterpart for the best fit values of the parameters as given by table \eqref{bestfit}. The shaded area represents the 1$\sigma$ confidence level.}
\end{figure}

The corner plot for the values of the parameters $H_0$,$H^\prime_0$, $h^{\prime\prime}_0$, $\Omega_{m0}$, $\xi$ and $\tilde{\sigma}$ with their $1\sigma$ and $2\sigma$ confidence levels is shown in figure \eqref{cornerplot}.
\subsection{The cosmological results}
The redshift evolution of the Hubble function and of the deceleration parameter $q$ are represented, for this model, in figure \eqref{fighubq}.
One can see from the figures that the behavior of the hybrid metric-Palatini model is very similar to the $\Lambda$CDM model. However, there are significant differences between the two models. As one can see from the figure, the Hubble parameter at present day is the same as its $\Lambda$CDM counterpart. The value of the Hubble parameter will become smaller than the $\Lambda$CDM value at redshifts $z\approx0.2$ and at earlier times the Hybrid metric-Palatini Hubble parameter is less than the $\Lambda$CDM value. This behavior can also be seen more quantitatively in diagrams for higher order derivatives of the Hubble parameter. The Taylor series expansion of the scale factor can be generally represented as \cite{SJ}
\begin{align}
	a(t)&=a_0\Big[1+H_0\left(t-t_0\right)-\frac{1}{2!}q_0H_0^2\left(t-t_0\right)^2\nonumber\\
	&+\frac{1}{3!}j_0H_0^3\left(t-t_0\right)^3+\frac{1}{4!}s_0H_0^4\left(t-t_0\right)^4+\mathcal{O}(5)\Big].
\end{align}
Based on this expansion one can introduce the jerk and snap parameters, defined as
\begin{align}\label{jands}
	j=\frac{1}{H^3}\frac{1}{a}\frac{d^3a}{dt^3},\qquad s=\frac{1}{H^4}\frac{1}{a}\frac{d^4a}{dt^4}.
\end{align}
In terms of the deceleration parameter $j$ and $s$ can be obtained as
\begin{align}
	j&=q+2q^2+(1+z)\frac{dq}{dz},\\
	s&=-(1+z)\frac{dj}{dz}-2j-3jq.
\end{align}
In figure \eqref{figjands} we have plotted the evolution of the jerk and snap parameters as a function of redshift. One can see from these figures that the behavior of the scale factor at higher derivatives is very different from the $\Lambda$CDM  model. The jerk parameter, as a convexity of the Hubble parameter, is constant in the $\Lambda$CDM  model. However we can see that its value varies in hybrid metric-Palatini model. In fact there is a range $z\in(0.5,2)$ where the jerk parameter is higher that its $\Lambda$CDM  counterpart. This is the region where the value change in Hubble parameter occurs. Also the snap parameter in $\Lambda$CDM  is a monotonically decreasing function of the redshift but its value in hybrid metric-Palatini model changes the slope at redshifts $z\approx1.5$. Since the snap parameter can be seen as a concavity of the deceleration parameter, one can see that the deceleration parameter becomes more concave at redshifts $z\gtrsim1.5$, making $q$ to becomes smaller that the $\Lambda$CDM  value at earlier times. This can be seen from figure \eqref{fighubq}. We see that the value of the deceleration parameter is greater than its $\Lambda$CDM  counterpart at present time, meaning that the hybrid metric-Palatini model predicts more acceleration for the present universe. However, as we have pointed out, because of the snap behavior, the deceleration parameter becomes less than $\Lambda$CDM  value at earlier times signaling less deceleration rate for early universe. In figures \eqref{figversus} we have also plotted the behavior of the jerk as a function of deceleration parameter and the snap as a function of jerk parameter. Although both of these plots are constants in $\Lambda$CDM  model, the hybrid metric-Palatini model predicts different behavior for them, verifying the above arguments.

The differences between the hybrid metric-Palatini model and the standard $\Lambda$CDM  model can also be seen through the values of the reduced chi-squared function for the Hubble function, which are shown in table~\ref{table1}. Also, in figure \eqref{hdiff} we have plotted this difference as a function of redshift. The shaded area represents 1$\sigma$ error. One can see from this figure that the hybrid metric-Palatini model predicts lower values of the Hubble parameter almost at all redshifts.
\begin{table}[h!]
	\centering
	\begin{tabular}{|c|c|}
		\hline
		Model & $\chi_{red}^2$  \\\hline\hline
		$\Lambda$CDM  & 1.051 \\\hline
		Hybrid metric-Palatini  & 1.056  \\\hline
	\end{tabular}
	\caption{The reduced $\chi^2$ for the Hubble function for the hybrid metric-Palatini gravity and the $\Lambda$CDM  model.}\label{table1}
\end{table}
The fluid nature of the effective part of the theory can also be investigated by considering the effective equation of state parameter defined as
\begin{align}
	\omega_{eff}=\frac{p_{eff}}{\rho_{eff}}.
\end{align}
\begin{figure}[h!]
	\includegraphics[scale=0.47]{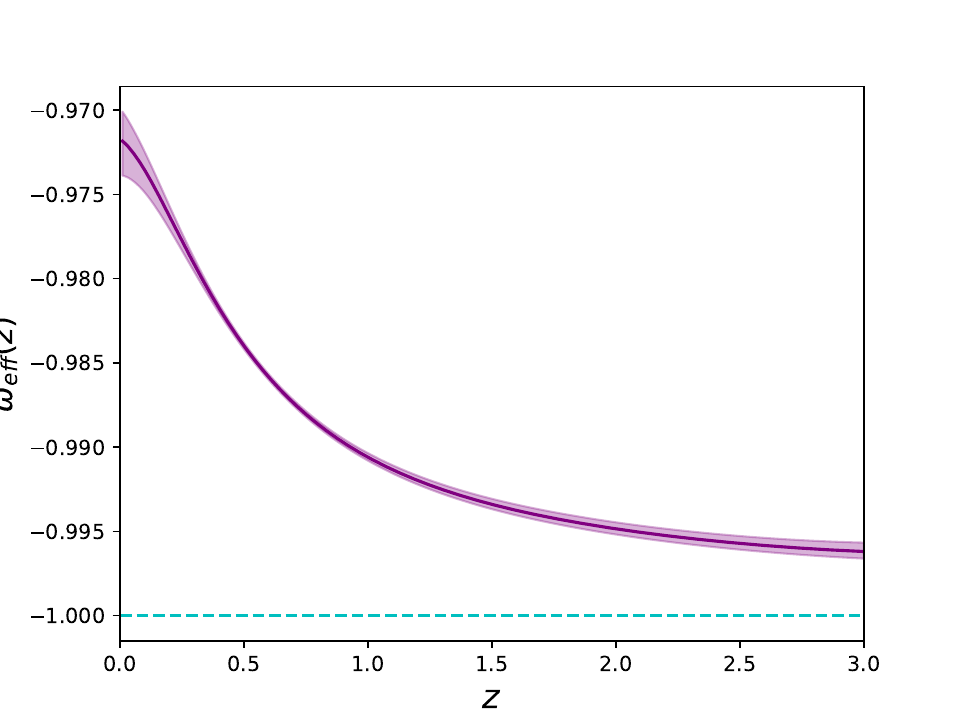}
	\caption{\label{figomega} The evolution of the effective equation of state parameter $\omega_{eff}$ as a function of redshift for the metric-Palatini $\sigma\mathcal{R}+\xi\mathcal{R}_{\mu\nu}\mathcal{R}^{\mu\nu}$ gravity model for the best fit values of the parameters as given by table \eqref{bestfit}. The shaded area denotes the $1\sigma$ error. Dashed lines represent $\Lambda$CDM model.}
\end{figure}
In figure \eqref{figomega} we have plotted the behavior the effective equation of state parameter as a function of redshift. One can see from the figure that the equation of state parameter is approximately constant and equal to the value of $\Lambda$CDM  model. More precisely, in hybrid metric-Palatini model, $\omega_{eff}$ is a decreasing function of redshift. The value of the equation of state parameter is a little larger than that of the $\Lambda$CDM  value signaling a slightly weaker acceleration. However, the value will become closer to the $\Lambda$CDM  value for earlier times.
\subsection{The statefinder diagnostic}
A very similar analysis could be done in the context of statefinder diagnosis, introduced in \cite{statefinder}. In this analysis we will use a pair $\{j,\bar{s}\}$ where $j$ is the jerk parameter defined in \eqref{jands} and $\bar s$ is a combination of the deceleration parameter $q$ as $j$ defined as
\begin{align}
	\bar{s}=\frac{j-1}{3(q-\frac{1}{2})}.
\end{align}
For $\Lambda$CDM model, the statefinder pair denotes a point $(1,0)$ in $(j,\bar{s})$ plane. 

\begin{figure}[h!]
	\includegraphics[scale=0.47]{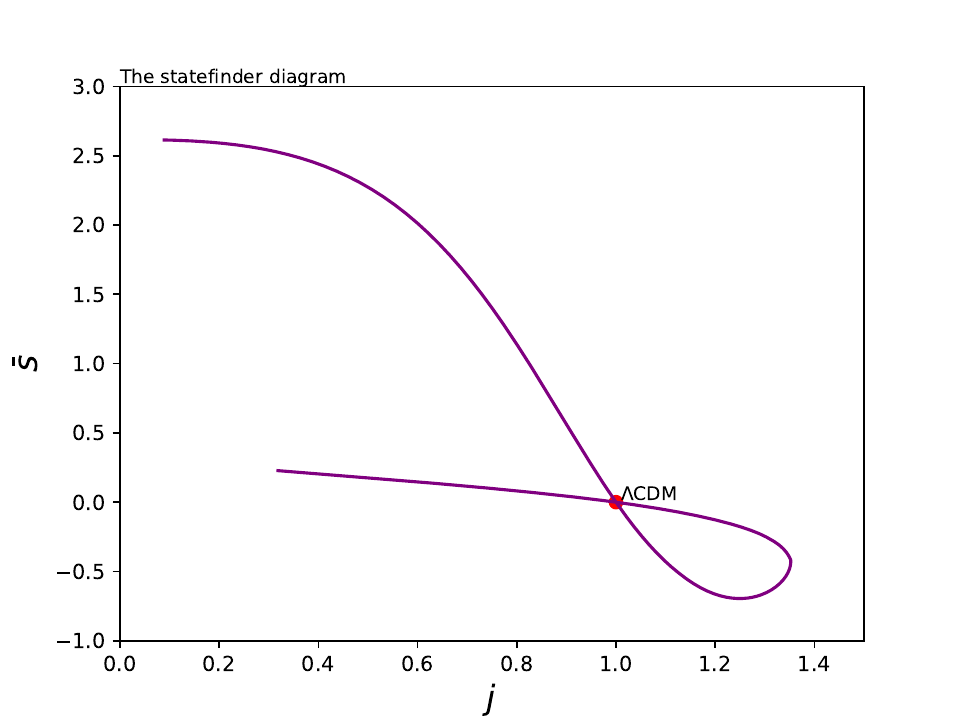}
	\caption{\label{statefinder} The variation of the statefinder variable $\bar{s}$ as a function of the jerk parameter $j$ for the metric-Palatini $\sigma\mathcal{R}+\xi\mathcal{R}_{\mu\nu}\mathcal{R}^{\mu\nu}$ gravity model for the best fit values of the parameters as given by table \eqref{bestfit}. Red dot represents $\Lambda$CDM model.}
\end{figure}

In figure \eqref{statefinder} we have depicted the variation of $\bar{s}$ as a function of $j$ in the metric-Palatini $\sigma\mathcal{R}+\xi\mathcal{R}_{\mu\nu}\mathcal{R}^{\mu\nu}$ model.
One can see from the figure that in the hybrid metric-Palatini model the state of the universe is not a point in the $(j,\bar s)$-plane. Actually, this is a general result of a modified gravity theory since the jerk parameters vary with redshift and differs from $+1$ as in $\Lambda$CDM model. However, the $\Lambda$CDM point seems to be an important point also in hybrid metric-Palatini model which is worth further study. For this paper, we just pointed out that generally the values $j<1$ and $\bar s>0$ indicates quintessence-like behavior for the dark energy sector but $j>1$ and $\bar s<0$ indicate phantom-like behavior. For the hybrid metric-Palatini model one can see that both cases happens and we have a quintessence and also phantom like behaviors in the evolution history of the universe. It should be noted that at present time the quintessence-like behavior is dominant . This can also be seen from the evolution of jerk parameter in figure \eqref{figjands} where the jerk parameter exceeds unity  for a redshift range $z\in(0.7,1.9)$. A final note worth pointing out in this section. As we have seen in before the equation of state parameter has a property $\omega_{eff}>-1$ which is the standard behavior of quintessence. However, we have seen that the phantom-like behavior can also be achieved in this model. This occurs normally in modified gravity theories, like k-essence models \cite{phantom} allowing to have a phantom behavior even with $\omega_{eff}>-1$ \cite{phantombehavior}. In our case, this can be related to the addition of the Palatini Ricci tensor in the h-metric tensor \eqref{hmetric}.
\subsection{The $Om(z)$ diagnostic} The $Om(z)$ diagnostic tool is an important theoretical method that allows to  distinguish alternative cosmological models from the standard  $\Lambda$CDM model. The $Om(z)$ diagnostic can be used to determine the nature of the dark energy fluid, and one could infer if the cosmological fluid  is a phantom-like fluid, a quintessence-like one, or it can be described by a simple cosmological constant. 
\begin{figure}[h!]
	\centering
	\includegraphics[scale=0.47]{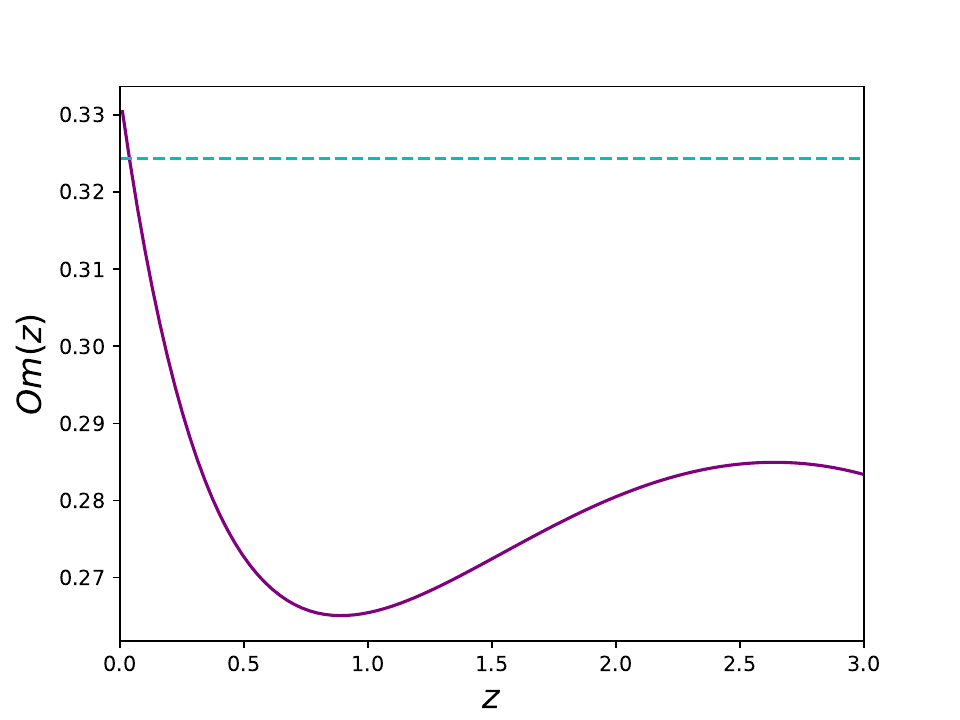}
	\caption{\label{Omdiag} The redshift variation of the $Om(z)$ diagnostic function for the Weyl-Boundary geometric gravity model for the best fit values of the parameters as given by table \eqref{bestfit}. The dashed line represents the $\Lambda$CDM model.}
\end{figure}
The $Om(z)$ function is defined as \cite{Om}
\begin{equation}\label{om}
	Om(z)=\frac{\left( H(z)/H_0\right)^{2}-1}{(1+z)^{3}-1}.
\end{equation}
For the standard $\Lambda$CDM model, the function $Om(z)$ is  equal to the present day matter density parameter $\Omega_{m0}$. In the case of cosmological models with a constant parameter of the equation of state of dark energy, $w=const.$, a positive slope of $Om(z)$ indicates  a phantom behavior, while a negative slope points towards  a quintessence-like evolution.  The evolution of the $Om(z)$ function as a function of redshift is plotted in figure \eqref{Omdiag}. One can see from the figure that the hybrid metric-Palatini effective fluid behaves as a quintessence at present times and also for earlier times $z>2$. However, for the redshift range $z\in(0.7,1.9)$ the slope of the diagram becomes positive and the effective fluid has a phantom-like behavior. This is in agreement with our previous discussions for the statefinder diagnostics. Also, one should note that the $Om$ values are almost always lower than the present day matter density abundance. This means that the effective energy density has a negative contribution to the energy budget of the universe.
\subsection{Variable equation of state alternative}
In this subsection we will obtain a suitable equation of state parameter of dark energy by fitting the resulting Hubble diagram with the predictions of hybrid metric-Palatini presented in figure \eqref{fighubq}.
\begin{figure}[h!]
	\centering
	\includegraphics[scale=0.47]{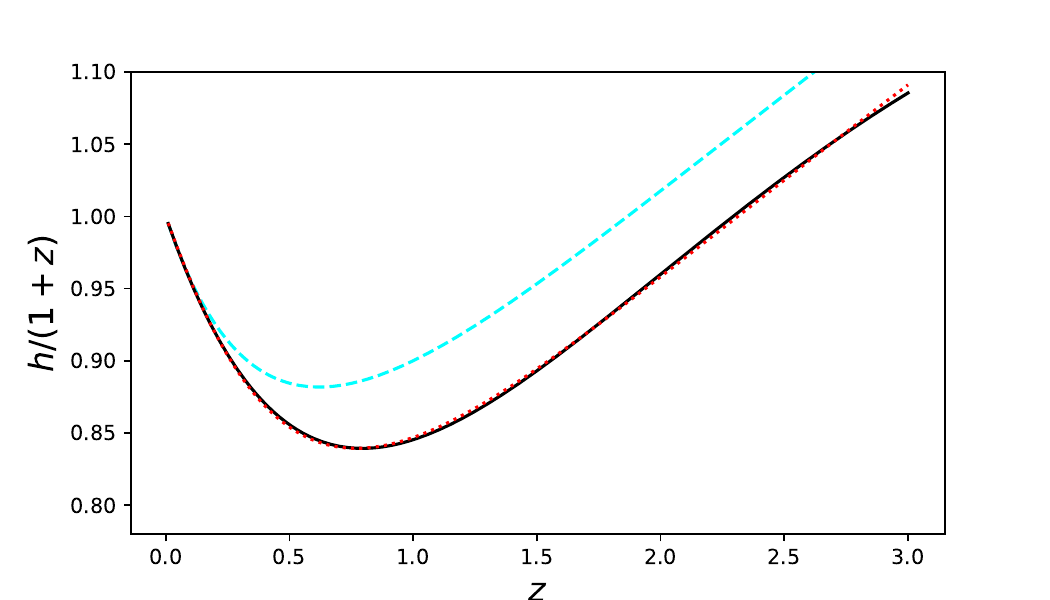}
	\caption{\label{hvswcdm} The redshift variation of the Hubble parameter for the hybrid metric-Palatini model (solid line) and the $\omega$-variable fitted model (dotted line) we the fitted values of the parameters as given by table \eqref{tab2}. The dashed line represents the $\Lambda$CDM model.}
\end{figure}
For this, let us assume that the accelerated expansion of the universe can be described by a exotic fluid with variable equation of state parameter of the form $p=\omega(z)\rho$. In this case The Hubble parameter takes the form
\begin{align}
	h^2=\Omega_{m 0}(1+z)^3+(1-\Omega_{m 0})e^{3\int_0^z\frac{1+\omega(z^\prime)}{1+z^\prime}dz^\prime}.
\end{align}
Phenomenologically, the dynamical dark energy could be well explained with CPL equation of state of the form \cite{CPL}
\begin{align}
	\omega(z)=w_0+w_a\frac{z}{1+z},
\end{align}
where $w_0$ and $w_a$ are arbitrary constants. In this case one obtains
\begin{align}\label{variableh}
	h^2&=\Omega_{m0}(1+z)^3\nonumber\\&+(1-\Omega_{m0} ) e^{-\frac{3 w_a z}{z+1}} (z+1)^{3 (w_0+w_a+1)}.
\end{align}
Now, we will obtain the constants $w_0$, $w_a$ and $\Omega_{m0}$ by fitting the model \eqref{variableh} with the prediction of the hybrid metric-Palatini model depicted in figure \eqref{fighubq}. The result is summarized in table \eqref{tab2}.
\begin{table}[h!]
	\centering
	\begin{tabular}{|c|c|}
		\hline
		Parameter & Fitted value  \\\hline\hline
		$\Omega_{m0}$  & 0.293 \\\hline
		$w_0$  & -0.926  \\\hline
		$w_a$ & -0.797\\\hline
	\end{tabular}
	\caption{The fitted values of the variable-$\omega$ model parameters confronting to the hybrid metric-Palatini prediction.}\label{tab2}
\end{table}
In figure \eqref{hvswcdm} we have plotted the hybrid metric-Palatini prediction (solid line), the equivalent variable-$\omega$ model (dotted line) and also the standard $\Lambda$CDM prediction (dashed line). One can see that the fitted parameter is in a very good agreement with the hybrid metric-Palatini model. As a result, one can see that at least with respect to the Hubble parameter, the hybrid metric-Palatini model can be described by CPL equation of state parameter.
\section{Conclusions and final remarks}
In this paper, we will consider the cosmological implications of a generalized hybrid metric-Palatini gravity. The hybrid metric-Palatini models are constructed in such a way that two different structures coupled in a gravitational action of the theory, one is constructed solely by the metric tensor; the connection being the Christoffel symbol, and the other is constructed by a metric and an independent affine connection. In this paper, we have considered the coupling of the form $f(\mathcal{R},\mathcal{R}_{\mu\nu}\mathcal{R}^{\mu\nu})$ in which the Palatini Ricci tensor appears explicitly in the action. Normally, in hybrid metric-Palatini models, the theory can be written by some field redefinitions, as a scalar-tensor theory of gravity; the affine connection can be obtained from a new metric tensor which is compatible with the independent connection. This is similar to the procedure one has taken in Palatini approach. However, in our case, due to the term $\mathcal{R}_{\mu\nu}\mathcal{R}^{\mu\nu}$, one can not repeat the aforementioned procedure analytically. Fortunately, for maximally symmetric space-times one can use another method to obtain the new metric as a function of the components of the original metric tensor $g_{\mu\nu}$. In this paper, we have considered this method for the case of an FRW universe filled with standard perfect fluid matter field.

For the resulting cosmological model, we have considered the background implications by confronting the model with the observational data on the Hubble parameter. We have used the MCMC method for the fitting, with Gaussian prior for the cosmological parameters and flat prior for the model parameters.
We have seen that the fitted hybrid metric-Palatini model is in a very good agreement with the $\Lambda$CDM predictions. The Hubble parameter is almost the same as its $\Lambda$CDM counterpart. However, slight deviations can be observed for small redshifts, where the hybrid metric-Palatini model predicts larger values, implying that the hybrid metric-Palatini model is younger than $\Lambda$CDM model. From the deceleration parameter, one can see that we have a smaller acceleration rate of the hybrid metric-Palatini model  than $\Lambda$CDM model in the present time and also smaller deceleration rate at earlier times. More details can be found from the model by considering higher derivatives of the Hubble parameter. We have plotted the evolution of jerk and snap parameters as a function of redshift, which indicates significant deviations from the $\Lambda$CDM model. Although the jerk parameter is constant in the $\Lambda$CDM model, we have seen that it varies for the hybrid metric-Palatini model. Observing that the jerk parameter can indicate the convexity of the Hubble diagram, one can infer that the convexity of the hybrid metric-Palatini Hubble parameter is the same as $\Lambda$CDM model, but the intensity varies in time. The snap parameter in the $\Lambda$CDM model is a decreasing function of the redshift, but the slope changes for the hybrid metric-Palatini model at redshifts about $z\approx1.5$. This could explain the change in the convexity of the deceleration parameter as was shown in figure \eqref{fighubq}. 


The equation of state parameter of the effective fluid, indicated that it is very close to the cosmological constant value. However, its value deviates more from $-1$ at present times, helping the model to explain the observational data. The higher derivative diagnostics can also be done through the statefinder method where the evolution of the jerk parameter is obtained as a function of new snap parameter which is a combination of the deceleration and jerk parameters. The diagram can distinguish between quintessence and phantom behavior of the effective dark energy fluid. We have seen that the hybrid metric-Palatini model can possess both natures in the evolution history of the universe, being quintessence-like in the present time.
The fluid behavior can also be obtained from the recently developed $Om$ diagnostics. We have seen that the $Om$ function for the hybrid metric-Palatini model, in agreement with the statefinder analysis, suggests that the effective dark energy fluid in the model has a quintessence-like behavior at present time. 
At the end of the cosmological considerations, we have find the alternative $\omega$-varying model that fits the behavior of the hybrid metric-Palatini model at the background level. This alternative model will deviate the hybrid metric-Palatini predictions both in perturbative and also higher order corrections of the Hubble parameter both can be used as a good fluid-based alternative for the geometric-based hybrid metric-Palatini model.


\begin{thebibliography}{99}
\bibitem{super} A. G. Riess, et al., Astron. J.  116,  1009 (1998); S. Perlmutter, et al., Astrophys. J.  517,  565 (1999).
\bibitem{BAO} D. J. Eisenstein, et al., Astrophys. J.  633,  560 (2005).
\bibitem{firsteinstein} A. Einstein, Sitzungsber. Preuss. Akad. Wiss. Phys. Math. Kl. 1917, 142 (1917).
\bibitem{LCDM} P. A. R. Ade, et al., Astron. Astrophys.  594, A13 (2016).
\bibitem{CCproblem} S. Weinberg, Rev. Mod. Phys.  61, 1 (1989); J. Martin, Comptes Rendus Physique  13, 566 (2012).  
\bibitem{coniprob} P. J. Steinhardt, Critical Problems in Physics, Princeton University Press, Princeton (1997); H. E. S. Velten, R. F. vom Marttens, W. Zimdahl, Eur. Phys. J. C 74, 3160  (2014). 
\bibitem{tensions} E. Abdalla, et. al, JHEAP 2204, 002 (2022); J. Hu, F. Wang, Universe 2023, 9(2), 94.
\bibitem{fR} T. P. Sotiriou and V. Faraoni, Rev. Mod. Phys.  82, 451 (2010); S. Capozziello and M. De Laurentis, Phys. Rep.  509, 167 (2011); L. Amendola and S. Tsujikawa, Dark Energy: Theory and Observations, Cambridge University Press, 2010.
\bibitem{brane} L. Randall and R. Sundrum, Phys. Rev. Lett.  83, 4690 (1999); L. Randall and R. Sundrum, Phys. Rev. Lett.  83, 3370 (1999); N. Arkani-Hamed, S. Dimopoulos, and G. Dvali, Phys. Lett. B  429, 263 (1998).
\bibitem{massive} M. Fierz and W. Pauli, Proc. Roy. Soc. A  173, 211 (1939); S. F. Hassan, R. Rosen and A. Schmidt-May , JHEP 2012, 26 (2012); C. de Rham, G. Gabadadze, and A. J. Tolley, Phys. Rev. Lett.  106, 231101 (2011); S. Shahidi, Proc. 14th Marcel Grossmann Meeting, 2017; Z. Haghani, H. R. Sepangi, and S. Shahidi, Phys. Rev. D  87, 124014 (2013); N. Khosravi, H. R. Sepangi, and S. Shahidi, Phys. Rev. D  86, 043517 (2012); N. Khosravi, N. Rahmanpour, H. R. Sepangi, and S. Shahidi, Phys. Rev. D  85, 024049 (2012).
\bibitem{weyl} H. Weyl, Annalen der Physik  360, 117 (1918); T. Harko and S. Shahidi, Eur. Phys. J. C  84, 509 (2024); D.-I. Visa, T. Harko, and S. Shahidi, Phys. Dark Univ.  46, 101720 (2024); J.-Z. Yang, S. Shahidi, and T. Harko, Eur. Phys. J. C 82, 1171 (2022).
\bibitem{finsler} R. Hama, T. Harko, S. V. Sabau and S. Shahidi, Eur. Phys. J. C  81, 742 (2021); A. Bouali, H. Chaudhary, R. Hama, T. Harko, S. V. Sabau, M. San Martín, Eur. Phys. J. C  83, 121 (2023); R. Hama, T. Harko, S. V. Sabau, Eur. Phys. J. C  83, 1030 (2023).
\bibitem{ghost} D. G. Boulware and S. Deser, Phys. Rev. D  6, 3368–3382 (1972); M. Ostrogradsky, Mem. Acad. St. Petersbourg  6, 385 (1850); A. Klein and D. Roest, JHEP  07, 130 (2016).
\bibitem{scalarvector} J. W. Moffat, JCAP 2006, 004 (2006); G. W. Horndeski, Int. J. Theor. Phys. 10, 363 (1974); T. Kobayashi, Rep. Prog. Phys., 82, 086901 (2019); A. Nicolis, R. Rattazzi, and E. Trincherini, Phys. Rev. D 79, 064036 (2009); C. Deffayet, X. Gao, D. A. Steer, and G. Zahariade, Phys. Rev. D 84, 064039 (2011); G. Tasinato, JHEP, 2014, 067 (2014); L. Heisenberg, JCAP 2014, 015 (2014).
\bibitem{generalmatter} R. A. C. Cipriano, N. Ganiyeva, T. Harko, F. S. N. Lobo, M. A. S. Pinto, and J. L. Rosa, Universe 10, 339 (2024); M. Roshan and F. Shojai, Phys. Rev. D, 94, 044002 (2016); Ö. Akarsu, N. Katirci, and S. Kumar, PoS CORFU2017, 105 (2018).
\bibitem{derivative} Z. Haghani and S. Shahidi, Phys. Dark Univ. 30, 100683 (2020); P. Asimakis, S. Basilakos, A. Lymperis, M. Petronikolou, and E. N. Saridakis, Phys. Rev. D 107, 104006 (2023); Z. Haghani and S. Shahidi, Eur. Phys. J. Plus 135, 509 (2020).
\bibitem{secondderivativepaper} Z. Haghani, T. Harko, S. Shahidi, Phys. Dark Univ. 44, 101448 (2024).
\bibitem{fRT} T. Harko, F. S. N. Lobo, S. Nojiri, S. D. Odintsov, Phys. Rev. D  84, 024020 (2011).
\bibitem{fRLm} T. Harko, F. S. N. Lobo, Eur. Phys. J. C  70, 373 (2010).
\bibitem{fRTRT} Z. Haghani, T. Harko, F. S. N. Lobo, H. R. Sepangi, S. Shahidi, Phys. Rev. D  88, 044023 (2013).
\bibitem{fRTLm} Z. Haghani, T. Harko, Eur. Phys. J. C  81, 615 (2021).
\bibitem{mukohyama} O. Lacombe, S. Mukohyama, J. Seitz, JCAP  05, 064 (2024).
\bibitem{replytomukoohyama} T. Harko, M. A. S. Pinto, S. Shahidi, Phys. Dark Univ. 48, 101863  (2025).
\bibitem{palatini} A. Palatini, Rend. Circ. Mat. Palermo,  43, 203 (1919); A. Delhom, D. Rubiera-Garcia, Palatini Theories of Gravity and Cosmology, in: E. N. Saridakis et al. (eds.), Modified Gravity and Cosmology, Springer, Cham (2021).
\bibitem{generalPalatini} G. J. Olmo, Int. J. Mod. Phys. D  20, 413 (2011).
\bibitem{palatinipapers}S. Lee, Mod. Phys. Lett. A  23, 1388 (2008); J. Wu, G. Li, T. Harko, S. D. Liang, Eur. Phys. J. C  78, 430 (2018).
\bibitem{hybridmetricpalatini} T. Harko, T. S. Koivisto, F. S. N. Lobo and G. J. Olmo,  Phys. Rev. D 85, 084016 (2012); S. Capozziello, T. Harko, F. S. N. Lobo, G. J. Olmo, Universe 1, 199  (2015).
\bibitem{noether} D. Bak, D. Cangemi and R. Jackiw, Phys. Rev. D 49 (1994) 5173.
\bibitem{noetherpalatini} R. Dick, Int. J. theor. Phys. 32, 109 (1993); P. Wang, G. M. Kremer, D. S. M. Alves, X. Meng, Gen. Relativ. Gravit. 38 (2006) 517; 
\bibitem{generalizedHMP} N. Tamanini, C. G. Boehmer, Phys. Rev. D 87, 084031 (2013); R. Aliannejadi, Z. Haghani, IJAA 819, 1183 (2024).
\bibitem{cosHMP} B. Asfour, A. Bargach, Y. Ladghami, A. Errahmani, T. Ouali, Phys. Lett. B  853, 138679 (2024); J. L. Rosa, Eur. Phys. J. C  84, 895 (2024); I. D. Gialamas, A. Karam, T. D. Pappas, E. Tomberg, Int. J. Geom. Meth. Mod. Phys.  20, 2330007 (2023); M. He, Y. Mikura, Y. Tada, JCAP  05, 047 (2023); S. Capozziello, T. Harko, T. S. Koivisto, F. S.N. Lobo and G. J. Olmo, JCAP 04 (2013) 011.


\bibitem{BlackHMP} B. Asfour, A. Bargach, Y. Ladghami, A. Errahmani, T. Ouali, Phys. Dark Univ.  47, 101812 (2025); D. Demir, K. Gabriel, A. Kasem, S. Khalil, Phys. Dark Univ.  42, 101336 (2023); P. Dyadina, N. Avdeev, Eur. Phys. J. C  84, 103 (2024); K. A. Bronnikov, S. V. Bolokhov, M. V. Skvortsova, Grav. Cosmol.,  27, 358 (2021); C.-Y. Chen, Y.-H. Kung, P. Chen, Phys. Rev. D  102, 124033 (2020); N. Avdeev, P. Dyadina, S. Labazova, J. Exp. Theor. Phys.  131, 537 (2020); K. A. Bronnikov, S. V. Bolokhov, M. V. Skvortsova, Grav. Cosmol.  26, 212 (2020).
\bibitem{theoHMP} J. S. Gonçalves, A. F. Santos, EPL  136, 50002 (2021); A. Borowiec, A. Kozak, JCAP  07, 003 (2020); J. B. Jiménez, L. Heisenberg, T. S. Koivisto, JCAP  08, 039 (2018); M. V. dos Santos, J. S. Alcaniz, D. F. Mota, S. Capozziello, Phys. Rev. D  97, 104010 (2018).
\bibitem{CCdata} A. Favale, A. Gómez-Valent and M. Migliaccio, MNRAS  523, 3406 (2023).
\bibitem{PANdata} D. Brout \textit{et al.}, ApJ,  938, 110 (2022).
\bibitem{SJ}F. Y. Wang, Z. G. Dai, and S. Qi, A\& A 507, 53 (2009).
\bibitem{statefinder} V. Sahni, T. D. Saini, A. A. Starobinsky, U. Alam, JETP Lett. 77, 201 (2003).
\bibitem{phantom}C. Armendariz-Picon, V. Mukhanov, P. J. Steinhardt, Phys. Rev. D  63, 103510 (2001).
\bibitem{phantombehavior} Y. Cai, S. Capozziello, M. De Laurentis and E. N. Saridakis, Rept. Prog. Phys. 79, 106901 (2016).
\bibitem{Om}V. Sahni, A. Shafieloo, A. A. Starobinsky, Phys. Rev. D  78, 103502 (2008).
\bibitem{CPL} M. Chevallier and D. Polarski, Int. J. Mod. Phys. D  10, 213 (2001); E. V. Linder, Phys. Rev. Lett.  90, 091301 (2003).
\end{thebibliography}
\end{document}